\documentclass[pra,reprint,superscriptaddress,showpacs,amsmath, floatfix,amssymb,aps]{revtex4-1}
\makeatletter

\newcommand{\Rmnum}[1]{\expandafter\@slowromancap\romannumeral #1@}
\makeatother
\usepackage{graphicx}
\usepackage{subfigure}
\usepackage{dcolumn}
\usepackage{bm}
\usepackage{hyperref}
\usepackage{indentfirst}
\usepackage{braket}
\usepackage{amsmath}
\usepackage{ulem}
\usepackage{color}
\usepackage{float}
\hypersetup{colorlinks=true, citecolor=blue, urlcolor=blue, linkcolor=blue}
\begin{document}
\title{Delta-Learning approach combined with the cluster Gutzwiller approximation for strongly correlated bosonic systems }

\author{Zhi Lin}
\email{zhilin18@ahu.edu.cn}
\affiliation{School of Physics and optoelectronic engineering, Anhui University, Hefei 230601, China}
\affiliation{State Key Laboratory of Surface Physics and Department of Physics, Fudan University, Shanghai 200433, China}
\author{Tong Wang}
\affiliation{School of Physics and optoelectronic engineering, Anhui University, Hefei 230601, China}
\author{Sheng Yue}
\affiliation{School of Physics and optoelectronic engineering, Anhui University, Hefei 230601, China}

\begin{abstract}
The cluster Gutzwiller method is widely used to study the strongly correlated bosonic systems, owing to its  ability to provide a more precise description of quantum fluctuations. However, its utility is limited by the exponential increase in computational complexity as the cluster size grows. To overcome this limitation, we propose an artificial intelligence-based method known as $\Delta$-Learning. This approach constructs a predictive model by learning the discrepancies between lower-precision (small cluster sizes) and high-precision (large cluster sizes) implementations of the cluster Gutzwiller method,  requiring only a small number of training samples. Using this predictive model, we  can effectively forecast the outcomes of high-precision methods with high accuracy.  Applied to various Bose-Hubbard models, the $\Delta$-Learning method effectively predicts phase diagrams while significantly reducing the  computational resources and time. Furthermore, we have compared the predictive accuracy of $\Delta$-Learning with other direct learning methods and found that $\Delta$-Learning exhibits superior performance in scenarios with limited training data. Therefore, when combined with the cluster Gutzwiller approximation, the $\Delta$-Learning approach offers a computationally efficient and accurate method for studying phase transitions in  large, complex bosonic systems.
\end{abstract}
\maketitle
\section{Introduction}
Ultracold bosonic systems  have emerged as a  clean and flexible platform for simulating the intriguing quantum many-body physics which lack natural counterparts in condensed matter physics \cite{Lewenstein1,Bloch1}. In particular, the experimental realization of the Bose-Hubbard model in optical lattices and the corresponding observation of  the superfluid to Mott insulator phase transition \cite{Nature415,np1} provide a platform for studying the novel quantum phenomenon in strongly correlated bosonic systems. Inspired by the realization of the Bose-Hubbard model in cubic lattices, various interesting strongly correlated bosonic physics have been revealed in the experiments, such as the realization of two-component Bose-Hubbard model \cite{TCBHM} and inhomogeneous Bose-Hubbard model \cite{NC}, as well as the observations of spin-dependent superfluid-to-Mott-insulator transition \cite{hexagonal1}, multi-orbital superfluid phase \cite{hexagonal2},  dynamic localization \cite{dynamic-l1,dynamic-l2,dynamic-l3}, density-dependent tunneling \cite{density-dependent-hopping1,density-dependent-hopping2}, spin-Mott state \cite{Spin-mott}, and frustrated chiral dynamics \cite{chiral-dynamics}. To understand these novel strongly correlated bosonic phenomena theoretically, researchers  have introduced several analytical methods, such as mean-field theory \cite{SC1},  strong-coupling expansion (SCE) method \cite{SC1,SC2,SC3}, the generalized effective-potential Landau theory (GEPLT) \cite{wang,zhi-4,zhi-5,Yue}and bosonization \cite{Bosonization}. Additionally, they  have developed numerical methods, including quantum Monte Carlo (QMC)~\cite{QMC-TBHM,QMC_ED-inho,wang,QMC-chiral}, exact diagonalization (ED) \cite{QMC_ED-inho}, density-matrix renormalization group method (DMRG) \cite{DMRG-inho,DMRG-density-DHopp,DMRG-density-DHopp2,DMRG-chiral1,DMRG-chiral2}  as well as cluster Gutzwiller method \cite{Gutzwiller1,Gutzwiller2,hexagonal-Gutzwiller,Gutzwiller3,Gutzwiller4,Gutzwiller5,Gutzwiller6}.  Although those analytical methods have the potential to acquire the accurate locations of the boundaries, they have  limitations. Specifically, the GEPLT can achieve approaching  QMC predictions for second-order phase transition, but it is invalid for first-order phase transition \cite{zhi-4,zhi-5}, the SCE method is a non-universal theory \cite{zhi-5} and performs  worse than  GEPLT for obtaining the quantum phase diagrams of ultracold dipolar Bose gases \cite{zhi-5}, while bosonization method can only handle quasi-one-dimensional systems.  Numerical methods like QMC, ED and DMRG are  powerful tools for studying the strongly correlated bosonic physics, but they require substantial computational resources. Fortunately, the cluster Gutzwiller method  can yield results  close to QMC simulations, while only using significantly less computational power~\cite{Gutzwiller2}.

Compared to the single-site Gutzwiller method, the cluster Gutzwiller method effectively captures the correlation effects for  larger clusters. By replacing a single lattice site with a supercell, this method can more accurately describe quantum fluctuations, thus overcoming the limitations of the single-site mean-field method \cite{Gutzwiller2,Gutzwiller5,Gutzwiller6}. However, when we use cluster Gutzwiller method to perform calculations,  the computational resources  and time  will grow exponentially with increasing the sizes of the cluster \cite{Gutzwiller2,Gutzwiller5,Gutzwiller6}. In order to obtain the  high-precision phase diagrams,  larger clusters must be considered by using cluster Gutzwiller method. Therefore, it is significant to optimize the cluster Gutzwiller method for saving the computational resources and time.

Fortunately, in quantum chemistry, there is already a mature method, namely the $\Delta$-Learning, which can  save the computational time and  resources while achieving high-quality results  \cite{Delta-Learning1,Delta-Learning2,Delta-Learning}. Specifically, $\Delta$-Learning approach is a machine learning (ML) method that develops a high-precision predictive model using a small number of training samples by learning the differences between low-precision and  high-precision calculations. By integrating the results of the  predictive model  with those from the low-precision methods, it can effectively forecast the outcomes of high-precision methods. This kind of technique has been demonstrated to be effective in quantum chemistry.

In this paper, we demonstrate the application of the commonly used quantum chemistry method, $\Delta$-Learning, to acquire a high-precision prediction mode for estimating the phase boundaries of various different Bose-Hubbard models. Here, we use the results obtained from the cluster Gutzwiller method with small (large) sizes as the low-precision (high-precision) calculations within $\Delta$-Learning approach. Furthermore,  we have compared the predictions made by $ \Delta $-learning with those  produced by  direct learning. For a small number of training sets, we find that the predictions from $\Delta$-Learning (whether it is obtained based on the support vector machines (SVM) or backpropagation neural networks (BPNN)) are significantly better than direct learning.  Additionally, the predictions from  $\Delta$-Learning using SVM  outperform those from $\Delta$-Learning utilizing BPNN when the number of training data satisfies $n\le 4$. Using the SVM-based $\Delta$-Learning, we  can not only  predict the phase boundaries of various different Bose-Hubbard models with high-precision, but also  greatly minimize computation resources and time compared to the cluster Gutzwiller method, all while maintaining  the same level of accuracy.
﻿

\section{A brief review of the Gutzwiller method}
To ensure this paper is comprehensive, we will give a brief review of the Gutzwiller method.
 We start with the single-site Gutzwiller method, which assumes that the system wave function can be written as the product of single-site wave functions. For the Bose-Hubbard model, the single-site wave function for each lattice site $i$ can be expressed as $ \left| i  \right\rangle =  {\textstyle \sum_{n}} c_n \left| n  \right\rangle $ \cite{gutzwillerwavefunction1,gutzwillerwavefunction2,gutzwillerwavefunction6,gutzwillerwavefunction3,gutzwillerwavefunction4,gutzwillerwavefunction5} in the basis of Fock state $\left| n  \right\rangle$ with $n$ particles, and the corresponding expansion coefficients $c_n $ can be determined by the imaginary time evolution~\cite{gutzwillerwavefunction5}. However, it can also be determined through self-consistent diagonalization procedure as described below \cite{Gutzwiller2}. Specifically, the wave function of the entire system can be written as $\left| i  \right\rangle \left| \psi  \right\rangle $, where $ \left| \psi  \right\rangle $ is the wave function of all sites excluding site  $i$. Right now,  the conventional Hamiltonian can be expressed in the form
\begin{equation}
	\hat{H}=\hat{H}_{\psi}+\hat{H}_{i}+\hat{H}_{\psi i},
\end{equation}
where $ \hat{H}_{\psi} $ and $ \hat{H}_{i} $ operates solely on the corresponding subsystems. The $ \hat{H}_{\psi i} $ term describes the coupling between $ \left| i  \right\rangle $ and $ \left| \psi  \right\rangle $. Following normal logic, we need to know information about $ \left | \psi  \right\rangle $ in order to solve it self-consistently. So, we assume that we already know $ \left| \psi  \right\rangle $, then in the local Fock state basis $\{\left| n  \right\rangle\}$, the Hamiltonian matrix of the whole system reads
\begin{eqnarray}\label{HM}
	H_{mn}&=&\left\langle \psi \right| \hat{H}_{\psi} \left| \psi  \right\rangle \delta_{mn}+\left\langle n \right|\hat{H}_{i}\left|  m  \right\rangle \nonumber\\
	&+&\left\langle \psi \right| \left\langle n \right|\hat{H}_{\psi i}\left|  m  \right\rangle \left| \psi  \right\rangle,
\end{eqnarray}
where the first term represents a constant energy shift. So we can get the wave function on site $i$ by diagonalizing $ H_{mn} $ in the small Fock basis. For the Bose-Hubbard Hamiltonian
\begin{equation}\label{BH}
	\hat{H}_{BH}=-J\sum_{\left\langle i,j \right\rangle}\hat{b}^{\dagger}_{i}\hat{b}_{j}+\frac{U}{2}\sum_{i}\hat{n}_{i}(\hat{n}_{i}-1)-\mu\sum_{i}\hat{n}_{i},
\end{equation}
where $J$ is the nearest-neighbor tunneling, $ \hat{b}^{\dagger}_{i} $ ($ \hat{b}_{i} $) is bosonic creation (annihilation) operator on site $ i$. Here, $\hat{n}_{i}=\hat{b}^{\dagger}_{i}\hat{b}_{i}$ is the particle-number operator on site $i$, $ U $ denotes the on-site repulsion, $ \mu $ represents the chemical potential.  The uncoupled part of the Hamiltonian $ \langle n|\hat{H}_{i}\left|  m  \right\rangle$ reads $ \langle n|\hat{H}_{i}\left|  m  \right\rangle =\frac{1}{2}Un(n-1)\delta_{mn}-\mu n \delta_{mn} $, while  the coupling term $\left\langle \psi \right| \left\langle n \right|\hat{H}_{\psi i}\left|  m  \right\rangle \left| \psi  \right\rangle$ can write
\begin{equation}\label{HM1}
	\hat{H}_{\psi i}=-J\left(\hat{b}^{\dagger}_{i}\sum_{\left\langle j \right\rangle} \langle\hat{b}_{j}\rangle  + \hat{b}_{i}\sum_{\left\langle j \right\rangle} \langle\hat{b}^{\dagger}_{j} \rangle \right),
\end{equation}
where the $ \left\langle j \right\rangle $ requires the summation over all nearest neighbors of site $j$. It is easily to know that the mean field order $\langle\hat{b}_{j}\rangle $ can read $ \langle\hat{b}_{j}\rangle = \langle\psi|\hat{b}_{j} | \psi  \rangle =\sum_{n} c^{*}_{n} c_{n+1} \sqrt{n+1}$ by using the Fock coefficients $ c_n $ on site $ j $. Because a single lattice site is only coupled with the mean field $ \langle\hat{b}_{j}\rangle $ and its conjugate, the single-site Gutzwiller method is equivalent to the mean-field approximation for the Bose-Hubbard model. Through diagonalizing the Hamiltonian matrix $ H_{mn}$, a new  value of the mean field order parameter $ \langle\hat{b}\rangle =\langle\hat{b}_{j}\rangle $ on site $j$ can be determined. By employing a self-consistent iterative scheme, the Hamiltonian can be effectively solved without prior knowledge of the wave function $ | \psi \rangle $.

Similarity, the cluster Gutzwiller method can be constructed by extending the single-site Gutzwiller approximation \cite{hexagonal-Gutzwiller}. Generally, if we  replace the single-site wave function $| i \rangle $ with the supercell cluster wave function $ | S  \rangle $ with $s$ sites,  the wave function of the entire system can be written as $\left| S  \right\rangle \left| \psi  \right\rangle $. Here, in many-sites Fock space basis  $ \{ | N  \rangle \}=\{ |n_{0},n_{1},n_{2},\dots  \rangle \} $,  the  lowest eigenvector for a supercell cluster can be written as $ | S  \rangle=\sum_{N}C_{N}| N  \rangle $,  where $ n_{i}=0,1,2,\dots $ represents the number of particles occupying site $i$ within the cluster $S$.  Drawing an analogy with expressions Eq. (\ref{HM}) and Eq. (\ref{HM1}), the corresponding Hamiltonian matrix $\hat{H}_{MN}$ for the Bose-Hubbard model is expressed by
\begin{equation}\label{CHM1}
	 \hat{H}_{MN}\!=\!\left\langle M \right| \!\hat{H}^{S}_{BH}-J\!\!\sum_{i\in \partial S}\!\!v_{i}(\hat{b}^{\dagger}_{i} \langle\hat{b}\rangle  + \hat{b}_{i} \langle\hat{b}^{\dagger}\rangle )| N  \rangle ,
\end{equation}
where the energy shift $ \left\langle \psi \right| \hat{H}_{\psi} \left| \psi  \right\rangle $ has been neglected. Here, $ \hat{H}^{S}_{BH} $ is the Bose-Hubbard Hamiltonian (see Eq. (\ref{BH})), and the summations of the lattice sites in $ \hat{H}^{S}_{BH} $ are limited within the cluster $ S $. The second term represents the interaction of all sites located at the boundary $ \partial S $ of the cluster with the mean-field amplitude $\langle \hat{b} \rangle $. The coefficient $v_{i}$ quantifies the number of connections between the site and the mean field. As shown in Eq. (\ref{CHM1}), under the assumption of a uniform lattice,  wherein all boundary sites interact with a common mean field $\langle \hat{b} \rangle $. However, this assumption can be readily generalized to accommodate superlattices or finite lattice systems. By diagonalizing Hamiltonian matrix $ \hat{H}_{MN} $, we can obtain the lowest eigenvalue and the lowest eigenvector $ | S  \rangle=\sum_{N}C_{N}| N  \rangle  $ of the supercell cluster, which is used to calculate the mean field,
\begin{equation}
	\langle \hat{b} \rangle=\langle S | \hat{b} | S \rangle=\sum_{N,M}C^{*}_{M}C_{N}\langle M | \hat{b}| N \rangle.
\end{equation}
By employing a self-consistent loop, we can eventually obtain the self-consistent lowest eigenvector $ | S  \rangle=\sum_{N}C_{N}| N  \rangle $.

Cluster Gutzwiller method is an effective approach for studying the Bose-Hubbard model. However, as research questions become more complex,  larger cluster need to be considered to more accurately describe quantum fluctuations. This leads to an exponential increase in the dimensions of the Hilbert space, resulting in larger Hamiltonian matrix dimensions, which pose significant challenges for computer memory storage. Additionally, this expansion results in slower  computational speed, leading to inefficient utilization of  computational resources. Fortunately, the rapid development of ML technology provides effective methods to address this issue. We can use the cluster Gutzwiller method to calculate a limited number of points in the phase diagram as training sets, and then employ ML to train a predictive model that can forecast the complete phase transition boundaries.

\section{$ \Delta $-Learning for ultracold boson}
In the following, we will introduce a ML technology, namely $\Delta$-learning, which  effectively forecast  the outcomes of high-precision methods  while  using  minimal  computational resources and time. In quantum chemistry (QC), $\Delta$-learning can predict the results of high-level approaches  based on low-level QC  methods by learning the differences between low- and high-level QC  values \cite{Delta-Learning1,Delta-Learning}. This method primarily focuses on identifying these difference,  characterizing $\Delta$-learning  as a ML technique to enhance a low-level QC method.

\begin{figure}[h!]
	\centering
	\includegraphics[width=0.7\linewidth]{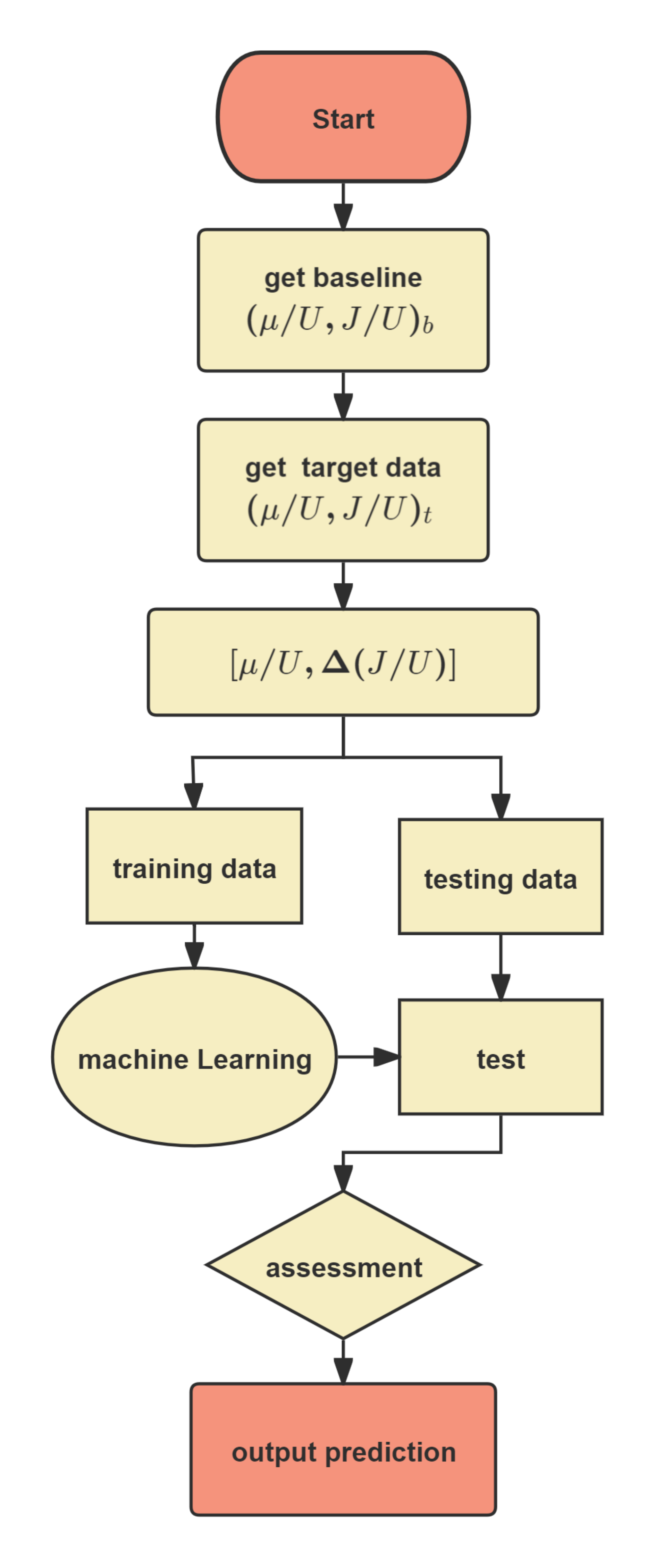}
	\caption{ Program flow chart of $\Delta$-learning.}
	\label{algorithm-1}
\end{figure}

To generate new predictions of the target property $y_t$ utilizing $\Delta$-learning, it is imperative to initially compute the baseline property $y_b$. Subsequently, the deviations $\Delta^{t}_{b}$ learned by the ML model are added to the baseline to yield the final prediction, as expressed by the following equation
\begin{equation}
	y_{t}=y_{b}+\Delta^{t}_{b}.
\end{equation}
In general, the computations of $\Delta^{t}_{b}$ using ML are substantially more rapider than the calculations of the baseline property $y_b$ via QC methods. Therefore, the computational expense of the $\Delta$-learning approach is predominantly governed by the cost of the calculations of the baseline. The program flowchart of $\Delta$-learning is shown in Fig.\ref{algorithm-1}.

To study the phase diagrams of ultracold bosonic systems,  $\Delta$-learning can be constructed by using  results  from the cluster Gutzwiller method with small (large) sizes as the low-precision (high-precision) calculations. As mentioned above,  when we implement $\Delta$-learning to predict the high-precision  phase diagrams, the calculations of the baseline using the cluster Gutzwiller method with small sizes are necessity, and then it leads to inevitable loss of computing resources. Here, we further compare the respective advantages of $\Delta$-learning and direct learning, where the baseline is not necessity. The disadvantage of $\Delta$-learning compared to direct learning is  in need of the more computing resources to calculate the baseline. However, there are two  significant advantages of $\Delta$-learning resulted from establishing the baseline: (1) we assume that the valves  from the baseline  serve as approximation to the target values and (2) in most of the cases,  the correctness of physical behavior can be ensured by the baseline method \cite{Delta-Learning1,Delta-Learning2,Delta-Learning}.

\begin{figure}[h!]
	\centering
	\includegraphics[width=1.0\linewidth]{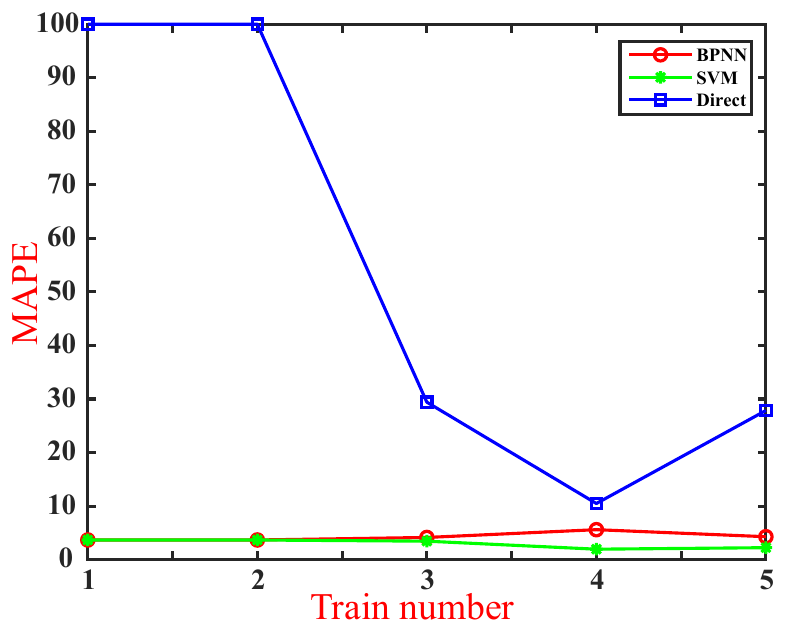}
	\caption{ Comparison of accuracy between direct learning and $\Delta$-learning. The horizontal axis represents the number of training samples, while the vertical axis denotes the mean absolute percentage error (MAPE) between the predicted and target  values. The blue square line indicates the variation of the MAPE with the number of training samples for the direct learning approach using support vector machines (SVM). The red circle (green asterisk) line shows the MAPE of the $\Delta$-learning model trained by BPNN (SVM) with various the number of training samples.
	}
	\label{Comparison}
\end{figure}
Obviously, larger training sets require  more computing resources to construct a predictive model. Therefore,  our goal is to develop  a high-accuracy learning method with smaller training sets. As shown in Fig.~\ref{Comparison}, $\Delta$-learning exhibits superior prediction accuracy compared to direct learning with fewer training samples. Additionally, by employing various ML models to implement the $\Delta$-learning, we can identify the optimal techniques for  executing the $\Delta$-learning. The results in Fig.~\ref{Comparison} indicate that in implementing the $\Delta$-learning,   the mean absolute percentage error (MAPE) from using SVM  is smaller than that from using BPNN when the number of training samples satisfies $n \ge3$.  Noteworthily, the MAPE from using SVM remains  very small for $n \ge4$. Consequently, we will utilize SVM to employ $\Delta$-Learning in various Bose-Hubbard models for achieving  the phase diagrams in the following section.

\section{$\Delta$-Learning approach for achieving phase diagrams in various Bose-Hubbard models }
\begin{figure}[h!]
	\centering
	\includegraphics[width=1.0\linewidth]{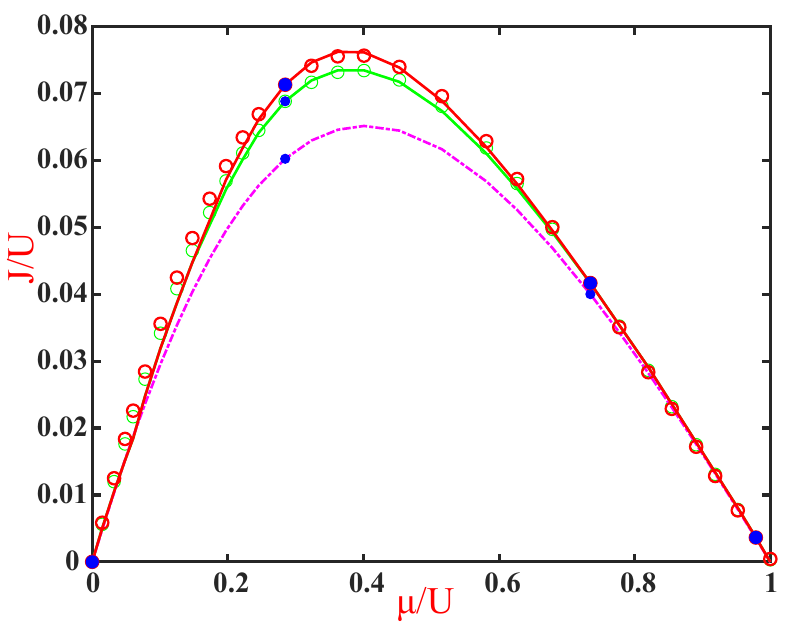}
	\caption{The predicted  phase boundaries of ultracold Bose gases in hexagonal optical lattice obtained via SVM-based $\Delta$-Learning. The green (red) circles represent the predicted values  obtained via SVM-based $\Delta$-Learning,  while the green (red) solid line describes results directly calculated by the cluster Gutzwiller method with $4\times3$ ($3\times6$) cluster. The pink dot-dashed line denotes baseline calculated by the cluster Gutzwiller method with $2\times2$, and the blue points are selected as the training samples. Here, the data for the Clustered Gutzwiller approach are sourced from the Ref.~\cite{Gutzwiller2}.}
	\label{2}
\end{figure}

Now we demonstrate that the SVM-based $\Delta$-Learning approach  is highly effective for obtaining the phase diagrams of ultracold boson in both simple optical lattices and superlattices. For simple optical lattices systems,  by examining ultracold bosonic atoms in two different types of lattices, square lattice and non-Bravais lattices like hexagonal lattices, we  show that the SVM-based $\Delta$-Learning approach can predict the high-precision phase boundaries with only four numbers of training samples. Furthermore, we reveal that this method is also valid for forecasting high-precision phase boundaries of the bosons in bipartite superlattices.
\begin{figure}[h!]
	\centering
	\includegraphics[width=1.0\linewidth]{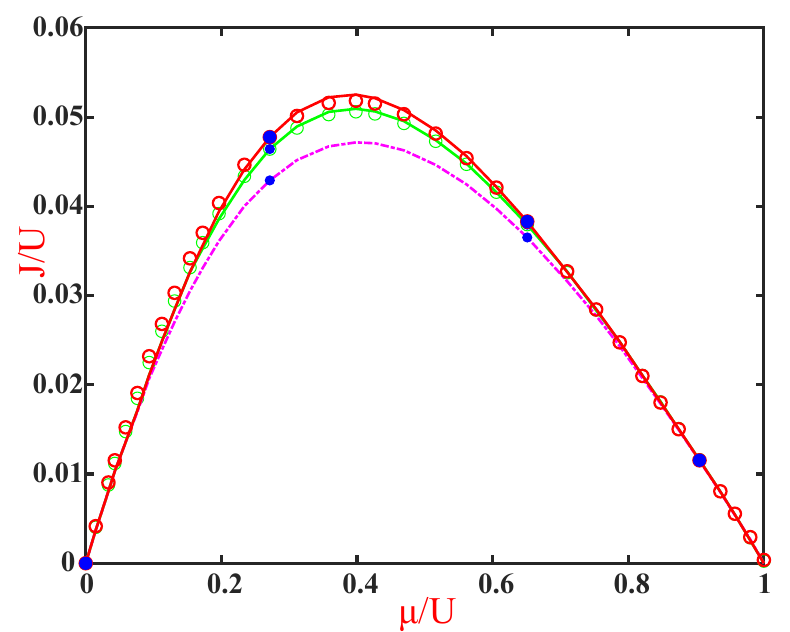}
	\caption{The predicted phase boundaries of ultracold Bose gases in square optical lattice gained through $\Delta$-Learning. The green (red) circles represent the predicted values  obtained through SVM-based $\Delta$-Learning, while the green (red) solid line describes results  directly calculated by the cluster Gutzwiller method with $3\times3$ ($4\times4$) cluster. The pink dot-dashed line represents baseline obtained by the cluster Gutzwiller method with $2\times2$,  and the blue points denote the training samples. Here, the data for the Clustered
Gutzwiller approach are originated from the Ref.~\cite{Gutzwiller2}.}
	\label{1}
\end{figure}
As shown in Fig.~\ref{1} the green (red) circles represent the predicted values  obtained via SVM-based $\Delta$-Learning, while  the green (red) solid line corresponds to the results  directly calculated by the cluster Gutzwiller method with $3\times3$ ($4\times4$) cluster.  The pink dot-dashed line serves as baseline, and the blue points are selected as the training samples. Firstly,  in the case of the square optical lattice,  we demonstrate the predicted  the phase boundaries for the standard Bose-Hubbard model utilizing the SVM-based $\Delta$-Learning. By comparing the values of the green (red) circles to the values of  the green (red) solid line in Fig.~\ref{1}, it is obvious  that they are in good agreement with each other. Furthermore, for hexagonal lattices systems, we also find that the predicted  phase boundaries obtained from the SVM-based $\Delta$-Learning are in excellent agreement with the results directly calculated by the cluster Gutzwiller method, as shown in Fig.~\ref{2}.

\begin{figure}[h!]
	\centering
	\includegraphics[width=1.0\linewidth]{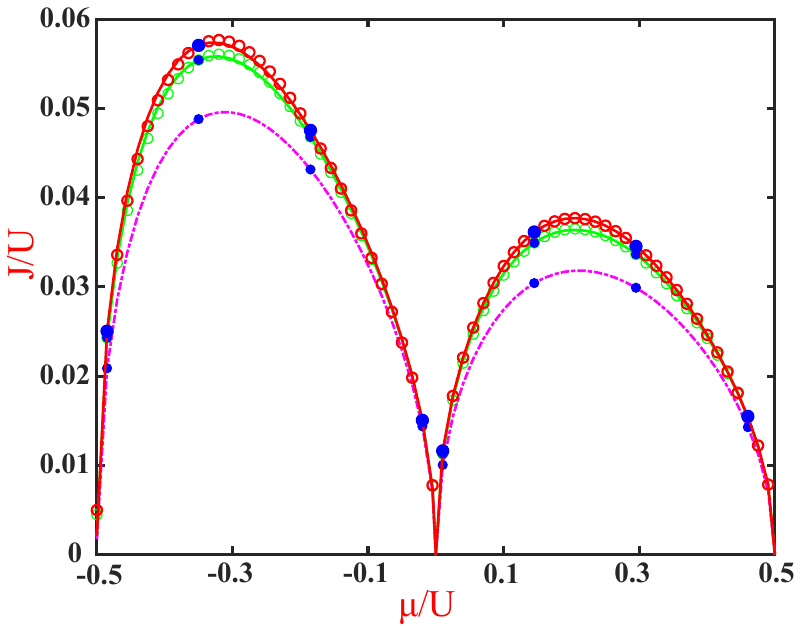}
	\caption{The predicted phase boundaries of ultracold Bose gases in bipartite superlattices  achieved via $\Delta$-Learning. The green (red) circles represent the predicted values  achieved via SVM-based $\Delta$-Learning, while the green (red) solid line describes results  directly calculated by the cluster Gutzwiller method with $2\times2$ ($4\times2$) cluster. The pink dot-dashed line signifies baseline obtained by the single-site Gutzwiller method, and the blue points are chosen as the training samples.}
	\label{3}
\end{figure}

\begin{figure}[h!]
	\centering
	\includegraphics[width=1.0\linewidth]{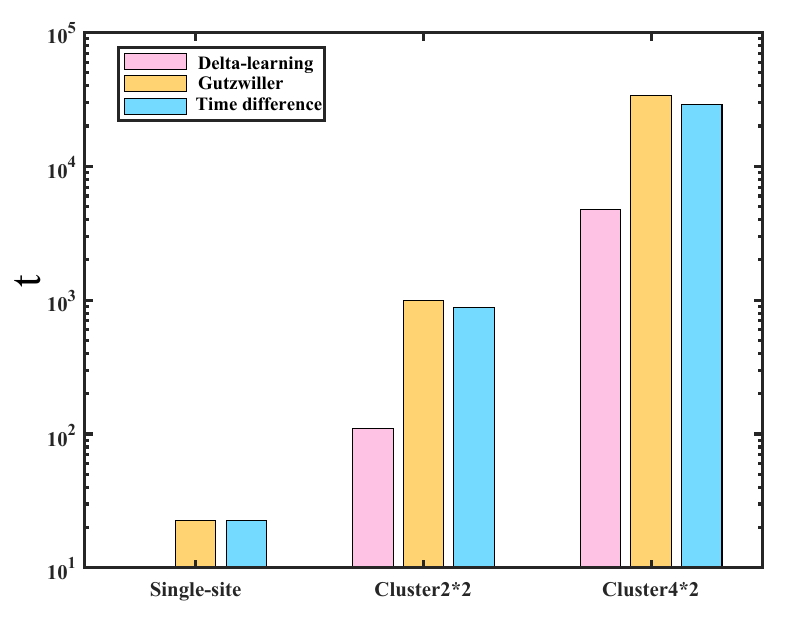}
	\caption{The computation time for achieving the phase boundaries presented in Fig.~\ref{3} by using  $\Delta$-Learning and Gutzwiller method with various clusters, respectively. The light coral represents the computation time by using $\Delta$-Learning,  the pale orange  denotes the computation time by using the various Gutzwiller methods, and the soft  cyan describes the computation time difference of those from two different methods. (CPU information : Intel(R) Core(TM) i9-14900KF   3.20 GHz ).}
	\label{4}
\end{figure}
Subsequently, we employ SVM-based $\Delta$-Learning to predicte the phase transition diagram of  ultracold boson in bipartite superlattices. The corresponding Hamiltonian of this  bipartite systems can be described  by
\begin{eqnarray}
	\hat{H}=&-&J\sum_{i\in A,j \in B
	}\hat{b}^{\dagger}_{i}\hat{b}_{j}+\frac{U}{2}\sum_{i\in A,i\in B}\hat{n}_{i}(\hat{n}_{i}-1)\nonumber\\
	&-&\mu\sum_{i\in B}\hat{n}_{i}-(\mu+\Delta \mu)\sum_{i\in A}\hat{n}_{i},
\end{eqnarray}
where $J$ represents the nearest-neighbor hopping parameter, $U$ is the strength of the on-site interaction, $b^{\dagger}_{i}$ is the boson creation operator, and $\Delta \mu$ is the chemical potential imbalance  between sublattices A and B. This model is valid for describing physics of  ultracold spinless bosonic atoms loaded into bipartite superlattices~\cite{superlattice1,NC,zhi-4}. Experimentally, the value of $\Delta \mu$ can be tuned by varying the parameters of the optical lattice. Without loss of generality, we set $\Delta \mu=0.5$ as example to
demonstrate  that the $\Delta$-Learning approach is highly effective for  predicting the phase diagrams of ultracold boson in bipartite superlattices with high-precision. As shown in Fig.~\ref{3},  we can easily find that the predictions from the SVM-based $\Delta$-Learning are in good agreement with the results directly calculated by cluster Gutzwiller method.

Furthermore, we also reveal the computation time for obtaining the phase boundaries presented in Fig.~\ref{3} by using  $\Delta$-Learning and Gutzwiller method with various clusters. As shown in Fig.~\ref{4},  the difference in computation time between $\Delta$-Learning approach  and cluster Gutzwiller method increases exponentially with the size of the cluster. It means that  the $\Delta$-Learning approach compared to the cluster Gutzwiller method can  significantly save the computational resources and time. Overall,  the $\Delta$-Learning method   is a numerically inexpensive approach for studying phase transitions in large, complex bosonic systems.

\section{Conclusion}
In this paper, we have proposed a $\Delta$-Learning approach  combined with the cluster Gutzwiller approximation  to study the phase diagrams of strongly correlated bosonic atoms in various optical lattices, including square, hexagonal, bipartite superlattices. Our results demonstrate that $\Delta$-Learning approach is a numerically inexpensive method that constructs a predictive model to effectively forecast the outcomes of the cluster Gutzwiller method for large cluster sizes. This is achieved by learning the difference between the cluster Gutzwiller results for small and large sizes  using only four training samples. Although the cluster Gutzwiller method can more accurately describe quantum fluctuations and is widely used to study the strongly correlated bosonic systems, it suffers from exponential growth in computational resources  and time  as cluster  size increases. Fortunately, by examining  bipartite superlattice systems, we reveal that the  $\Delta$-Learning method can significantly reduce the computational resources and time required. Therefore, the  $\Delta$-Learning approach combined  with  the cluster Gutzwiller approximation has tremendous potential especially  for further studies of phase transitions in  large, complex strongly correlated bosonic systems.

\section{Acknowledgements}\label{Acknowledgements}
The authors thank M. Yang, A. Ullah and  C. R. Liu for fruitful discussions.  This research is supported by the National Natural Science Foundation of China (NSFC) under Grant No.12004005, the Scientific Research Fund for Distinguished Young Scholars of the Education Department of Anhui Province No.2022AH020008, the Natural Science Foundation of Anhui Province under Grant No. 2008085QA26, and the open project of the state key laboratory of surface physics in Fudan University under Grant No. KF2021$\_$08.

\end{document}